# SOLUTION OF THE DIRAC EQUATION FOR POTENTIAL INTERACTION

A. D. Alhaidari
*Physics Department, King Fahd University of Petroleum & Minerals, Box 5047, Dhahran 31261, Saudi Arabia*
E-mail: haidari@mailaps.org

An effective approach for solving the three-dimensional Dirac equation for spherically symmetric local interactions, which we have introduced recently, is reviewed and consolidated. The merit of the approach is in producing Schrödinger-like equation for the spinor components that could simply be solved by correspondence with well-known exactly solvable non-relativistic problems. Taking the nonrelativistic limit will reproduce the nonrelativistic problem. The approach has been used successfully in establishing the relativistic extension of all classes of shape invariant potentials as well as other exactly solvable nonrelativistic problems. These include the Coulomb, Oscillator, Scarf, Pöschl-Teller, Woods-Saxon, etc.

## 1. Introduction

One may choose not to disagree with the view that exactly solvable systems are, by some (debatable) definitions, trivial. Nonetheless, exact solutions of the wave equation are important because of the conceptual understanding of physics that can only be brought about by such solutions. These solutions are also valuable means for checking and improving models and numerical methods being introduced for solving complicated physical problems. In fact, in some limiting cases or for some special circumstances they would constitute analytic solutions of realistic problems or approximations thereof. Most of the known exactly solvable problems fall within distinct classes of shape invariant potentials.[1] All potentials in a given class, along with their corresponding solutions (energy spectra and wavefunctions), can be mapped into one another by point canonical transformation (PCT).[2] Henceforth, only one problem (the "reference problem") in a given class needs to be solved to obtain solutions of all others in the class. PCT maintains the functional form of the wave equation (i.e. shape invariance of the potential). In other words, it leaves the canonical form of the wave equation invariant. As a result, a correspondence map among the potential parameters, angular momentum, and energy of the two problems (the reference and "target" problem) is obtained. Using this parameter substitution map and the energy spectrum of the reference problem one can easily and directly obtain the spectra of all other potentials in the class. Moreover, the wavefunctions are obtained by simple transformations of that of the reference problem. Supersymmetric quantum mechanics[3] and potential algebras[4] are two among other methods beside PCT which are used in the search for exact solutions of the wave equation. In nonrelativistic quantum mechanics, this development was carried out over the years by many authors where several classes of shape invariant potentials being accounted for and tabulated.[1] It was also extended to other classes of conditionally exactly[5] and quasi exactly[6] solvable problems where all or, respectively, part of the energy spectrum is known.

The relativistic extension of these formulations, on the other hand, remained for a long time only partially developed. Despite all the work that has been done over the years on the Dirac equation, its exact solutions for local interaction has been limited to a very small set of potentials. Since the original work of Dirac in the early part of last century up until 1989 only the relativistic Coulomb problem was solved exactly. In 1989, the relativistic extension of the oscillator problem (Dirac-Oscillator) was finally formulated and solved by Moshinsky and Szczepaniak.[7] Recently, and in a series of articles,[8-11] we presented an effective approach for solving the three dimensional Dirac equation for spherically symmetric potential interaction. Although the first steps were plagued with problems of misinterpretations,[12-14] the findings were always sound independent of those considerations. The first step in the program started with the realization that the nonrelativistic Coulomb, Oscillator, and S-wave Morse problems belong to the same class of shape invariant potentials which carries a representation of so(2,1) Lie algebra. Therefore, the fact that the relativistic version of the first two problems (Dirac-Coulomb and Dirac-Oscillator) were solved exactly makes the solution of the third, in principle, feasible. Indeed, the relativistic S-wave Dirac-Morse problem was formulated and solved in Ref. 8. The bound state energy spectrum and spinor wavefunctions were obtained. Taking the nonrelativistic limit reproduces the familiar S-wave Schrödinger-Morse problem. Motivated by these positive results, the same approach was applied successfully in obtaining solutions for the relativistic extension of yet another class of shape invariant potentials.[9] These included the Dirac-Scarf, Dirac-Rosen-Morse I & II, Dirac-Pöschl-Teller, and Dirac-Eckart potentials. Furthermore, using the same formalism quasi exactly solvable systems at rest mass energies were obtained for a large class of power-law relativistic potentials.[10] Quite recently, Guo Jian-You *et al* succeeded in constructing a solution for the relativistic Dirac-Woods-Saxon problem using the same approach.[15] In the fourth and last article of the series in our program of searching for exact solutions to the Dirac equation,[11] we found a special graded extension of so(2,1) Lie algebra. Realization of this superalgebra by 2×2 matrices of differential operators acting in the two component spinor space was constructed. The linear span of this graded algebra gives the canonical form of the radial Dirac Hamiltonian. It turned out that the Dirac-Oscillator class, which also includes the Dirac-Coulomb and Dirac-Morse, carries a representation of this supersymmetry. An "extended point canonical transformation" (XPCT) was introduced in Ref. 11. It maps all relativistic potentials belonging to a given class into one another. Therefore, just like in the nonrelativistic case, only one problem (the reference problem) in a given potential class needs to be solved to obtain solutions of all other relativistic problems in the class using the XPCT maps.

The approach is initiated by writing the relativistic Hamiltonian for a Dirac spinor coupled non-minimally to a four-component potential $(A_0,\vec{A})$. Spherical symmetry is imposed on the interaction by restricting the potential to the form $(A_0,\vec{A})$ = $[\lambda V(r),\hat{r}W(r)]$, where $\lambda$ is the Compton wavelength scale parameter $\hbar/mc$, and $\hat{r}$ is the radial unit vector. $V(r)$ and $W(r)$ are real radial functions referred to as the even and odd components of the relativistic potential, respectively. The resulting Dirac equation gives two coupled first order differential equations for the two radial spinor components. By eliminating one component we obtain a second order differential equation for the other. The resulting equation may turn out not to be Schrödinger-like, i.e. it may contain first order derivatives. A global unitary transformation is applied to the Dirac equation to eliminate the first order derivative. This, almost always, makes the solution of the relativistic problem easily attainable by simple and direct correspondence with well-known exactly solvable nonrelativistic problems. The correspondence results in a map among the relativistic and nonrelativistic parameters. Using this map and the known nonrelativistic energy spectrum one can easily and directly obtain the relativistic spectrum. Moreover, the two components of the spinor wavefunction are obtained from the nonrelativistic wavefunction using the same parameter map. The Schrödinger-like requirement produces a constraint in the form of a linear relation between the two potential components as $V \sim W + \kappa/r$, where $\kappa$ is the spin-orbit quantum number defined as $\kappa = \pm (j + ½)$ for $\ell = j \pm ½$. This will result in a Hamiltonian that will be written in terms of only one arbitrary potential function; either $V(r)$ or $W(r)$. Meeting the Schrödinger-like requirement is generally possible only because of the degree of flexibility brought about by the presence of two potential components. However, the unitary transformation is not necessary when $V = 0$. This corresponds to the case of the superpotentials $U^2 \pm U'$, where $U = W + \kappa/r$. The Dirac-Oscillator, where $W \sim r$, is an example of such a case. The Dirac-Scarf and Dirac-Pöschl-Teller potentials are also two among other such examples.

The paper is organized as follows. In Sec. 2, an overview of the technical details of the formalism will be presented. Implementation of the approach on selected potential examples will be given in the same section while a comprehensive list will be displayed in a tabular form. Sec. 3 will be devoted to a brief introduction to the extended point canonical transformation (XPCT), which will then be applied to the Dirac-Oscillator and Dirac-Rosen-Morse potential classes. The paper concludes with a short summary.

## 2. Overview of the Approach

In this section a brief summary will be given for the approach which is proposed to solve the three dimensional Dirac equation for spherically symmetric potential interactions. The aim is to produce Schrödinger-like equation for the spinor components making the solution of the relativistic problem easily obtainable by simple correspondence with well-known exactly solvable nonrelativistic problems.

In atomic units ($m = \hbar = 1$) and taking the speed of light $c = ƛ^{-1}$, we write the Hamiltonian for a Dirac particle coupled to a four-component potential $(A_0, \vec{A})$ as follows:

$$H = \begin{pmatrix} 1 + ƛA_0 & -iƛ\vec{\sigma}\cdot\vec{\nabla} + iƛ\vec{\sigma}\cdot\vec{A} \\ -iƛ\vec{\sigma}\cdot\vec{\nabla} - iƛ\vec{\sigma}\cdot\vec{A} & -1 + ƛA_0 \end{pmatrix}$$

where $\vec{\sigma}$ are the three hermitian 2×2 Pauli spin matrices. It should be noted that this type of coupling does not support an interpretation of $(A_0, \vec{A})$ as the electromagnetic potential unless, of course, $\vec{A} = 0$ (e.g., the Coulomb potential). Likewise, $H$ does not have local gauge symmetry. That is, the associated wave equation is not invariant under the usual electromagnetic gauge transformation. Imposing spherical symmetry and writing $(A_0, \vec{A})$ as $[ƛV(r), \hat{r}W(r)]$, we obtain the following Dirac equation

$$\begin{pmatrix} 1+\lambda^2 V(r)-\varepsilon & -i\lambda\vec{\sigma}\cdot\vec{\nabla}+i\lambda\vec{\sigma}\cdot\hat{r}W(r) \\ -i\lambda\vec{\sigma}\cdot\vec{\nabla}-i\lambda\vec{\sigma}\cdot\hat{r}W(r) & -1+\lambda^2 V(r)-\varepsilon \end{pmatrix}\begin{pmatrix} i[g(r)/r]\chi_{lm}^{j} \\ [f(r)/r]\vec{\sigma}\cdot\hat{r}\chi_{lm}^{j} \end{pmatrix}=0$$

where $\varepsilon$ is the relativistic energy and the angular wavefunction for the two-component spinor is written as[16]

$$\chi_{\ell m}^{j}(\Omega)=\frac{1}{\sqrt{2\ell+1}}\begin{pmatrix} \sqrt{\ell\pm m+1/2}\;\; Y_{\ell}^{m-1/2} \\ \pm\sqrt{\ell\mp m+1/2}\;\; Y_{\ell}^{m+1/2} \end{pmatrix}; \quad \text{for } j=\ell\pm ½$$

$Y_{\ell}^{m\pm 1/2}$ is the spherical harmonic function. Using spherical symmetry, which gives $\left(\vec{\sigma}\cdot\vec{L}\right)\Psi(r,\Omega)=-(1+\kappa)\Psi(r,\Omega)$, and the relations

$$(\vec{\sigma}\cdot\vec{\nabla})(\vec{\sigma}\cdot\hat{r})F(r)\chi_{\ell m}^{j}=\left(\frac{dF}{dr}+\frac{1-\kappa}{r}F\right)\chi_{\ell m}^{j}$$

$$(\vec{\sigma}\cdot\vec{\nabla})F(r)\chi_{\ell m}^{j}=\left(\frac{dF}{dr}+\frac{1+\kappa}{r}F\right)(\vec{\sigma}\cdot\hat{r})\chi_{\ell m}^{j}$$

we obtain the following two component radial Dirac equation

$$\begin{pmatrix} 1+\lambda^2 V(r)-\varepsilon & \lambda\left[\frac{\kappa}{r}+W(r)-\frac{d}{dr}\right] \\ \lambda\left[\frac{\kappa}{r}+W(r)+\frac{d}{dr}\right] & -1+\lambda^2 V(r)-\varepsilon \end{pmatrix}\begin{pmatrix} g(r) \\ f(r) \end{pmatrix}=0 \qquad (2.1)$$

It results in two coupled first order differential equations for the two radial spinor components. Eliminating one component in favor of the other gives a 2nd order differential equation. This will not be Schrödinger-like (i.e., it contains first order derivatives) unless $V=0$. To obtain Schrödinger-like equation in the general case we proceed as follows. A global unitary transformation $\mathcal{U}(\eta)=\exp(\frac{i}{2}\lambda\eta\sigma_2)$ is applied to the Dirac equation (2.1), where $\eta$ is a real constant parameter and $\sigma_2$ is the 2×2 Pauli spin matrix $\left(\begin{smallmatrix} 0 & -i \\ i & 0 \end{smallmatrix}\right)$. The Schrödinger-like requirement relates the two potential components by the linear constraint $V(r)=\xi\left[W(r)+\kappa/r\right]$, where $\xi$ is a real parameter. It also requires that $\sin(\lambda\,\eta)=\pm\lambda\xi$. This results in a Hamiltonian that will be written in terms of only one arbitrary potential function; either the even potential component $V(r)$ or the odd one $W(r)$. Therefore, one of two possibilities will be encountered. In the first, $\kappa$ will appear explicitly in the Hamiltonian along with the arbitrary potential component. The Dirac-Coulomb and Dirac-Oscillator are examples of such a case, which is solved for a general angular momentum.[8] In the second, the odd potential component will always be written as $W(r)=U(r)-\kappa/r$, where $U(r)$ is arbitrary and $\kappa$-independent. This will eliminate any reference to $\kappa$ from the Hamiltonian as clearly seen by substituting this expression of $W(r)$ in Eq. (2.1). S-wave relativistic problems including all of those in Refs. 8, 9, and 15 are of this type. The unitary transformation together with the potential constraint maps Eq. (2.1) into the following one, which we choose to write in terms of the odd potential component

$$\begin{pmatrix} C-\varepsilon+(1\pm 1)\lambda^2\xi\left(W+\frac{\kappa}{r}\right) & \lambda\left[\mp\xi+C\left(W+\frac{\kappa}{r}\right)-\frac{d}{dr}\right] \\ \lambda\left[\mp\xi+C\left(W+\frac{\kappa}{r}\right)+\frac{d}{dr}\right] & -C-\varepsilon+(1\mp 1)\lambda^2\xi\left(W+\frac{\kappa}{r}\right) \end{pmatrix}\begin{pmatrix} \phi^{+}(r) \\ \phi^{-}(r) \end{pmatrix}=0 \quad (2.2)$$

where $C = \cos(\lambda\,\eta)$ and $\left(\begin{smallmatrix}\phi^+\\ \phi^-\end{smallmatrix}\right) = \mathcal{U}\left(\begin{smallmatrix}g\\ f\end{smallmatrix}\right)$. This gives the following equation for one spinor component in terms of the other

$$\phi^{\mp}(r) = \frac{\lambda}{C \pm \varepsilon}\left[-\xi \pm C\left(W + \frac{\kappa}{r}\right) + \frac{d}{dr}\right]\phi^{\pm}(r) \qquad (2.3)$$

While, the resulting Schrödinger-like wave equation becomes

$$\left[-\frac{d^2}{dr^2} + C^2U^2 \mp C\frac{dU}{dr} + 2\xi\varepsilon U - \frac{\varepsilon^2 - 1}{\lambda^2}\right]\phi^{\pm}(r) = 0 \qquad (2.4)$$

where $U = W + \kappa/r$. In all relativistic problems that have been successfully tackled so far, Eq. (2.4) is solved by correspondence with well-known exactly solvable nonrelativistic problems. This, of course, results in a dramatic reduction in the efforts needed to construct the solution of the relativistic problem. This welcomed simplification in the solution of the relativistic problem could be explained by looking at the nonrelativistic limit ($\lambda \to 0$). The limit, gives the following

$$\xi = \pm\sin(\lambda\eta)/\lambda \approx \pm\eta\,,\ C \approx 1 - \lambda^2\eta^2\,,\ \varepsilon \approx 1 + \lambda^2 E$$

where $E$ is the nonrelativistic energy. Substituting in Eq. (2.3) shows that the lower spinor component $\phi^-(r)$ vanishes (assuming that the wavefunctions are differentiable). On the other hand, Eq. (2.4) reduces to the following nonrelativistic equation

$$\left[-\frac{d^2}{dr^2} + U^2 \mp \frac{dU}{dr} + 2\xi U - 2E\right]\psi(r) = 0 \qquad (2.5)$$

Hence, the functional form of the wave equation is maintained in the limit. Therefore, it is justifiable to conclude that a correspondence map would, most certainly, be obtained by comparing the nonrelativistic Eq. (2.5) with the relativistic Eq. (2.4). This correspondence results in a parameter map that relates the two problems. Now, if the nonrelativistic problem is exactly solvable, then using this parameter map and the known nonrelativistic energy spectrum one can easily obtain the relativistic spectrum. In fact, the relativistic extension of any known dynamical relationship in the nonrelativistic theory could easily be obtained by this correspondence map. The Green's function, which has a prime significance in the calculation of physical processes, is such an example.[17] Moreover, the spinor component wavefunction is also obtained from the nonrelativistic wavefunction using the same parameter map. It is to be noted that, typically, there exist only one nonrelativistic Schrödinger equation to be compared with the two relativistic equations for $\phi^{\pm}$. Thus, two independent parameter maps will be obtained corresponding to the two spinor components.

The implementation steps of the approach will now be illustrated by applying it on two selected examples. First, we consider the Dirac-Coulomb problem where $W(r) = 0$. The Schrödinger-like requirement gives the potential constraint that yields $V(r) = \xi\kappa/r \equiv Z/r$, where $Z$ is the spinor charge. This gives $\xi = Z/\kappa$ and relates the transformation parameter to the physical constants of the problem as $S \equiv \sin(\lambda\eta) = \pm\lambda Z/\kappa = \pm\alpha\mathbb{Z}/\kappa$, where $\alpha$ is the fine structure constant and $\mathbb{Z}$ is the dimensionless spinor charge in units of $e$. The wave equation (2.4), for the upper spinor component reads

$$\left[-\frac{d^2}{dr^2} + \frac{\gamma(\gamma+1)}{r^2} + 2\frac{Z\varepsilon}{r} - \frac{\varepsilon^2 - 1}{\lambda^2}\right]\phi^+(r) = 0 \qquad (2.6)$$

where $\gamma \equiv C\kappa = \sqrt{\kappa^2 - ƛ^2 Z^2} = \sqrt{\kappa^2 - \alpha^2 \mathbb{Z}^2}$ is the relativistic angular momentum. Comparing this equation with the nonrelativistic Schrödinger-Coulomb wave equation

$$\left[-\frac{d^2}{dr^2} + \frac{\ell(\ell+1)}{r^2} + 2\frac{Z}{r} - 2E\right]\psi(r) = 0 \tag{2.7}$$

gives, by correspondence, the following map between the parameters of the two problems:

$$\begin{aligned} &\ell \to \gamma \quad \text{or} \quad \ell \to -\gamma - 1 \\ &Z \to Z\varepsilon \\ &E \to (\varepsilon^2 - 1)/2ƛ^2 \end{aligned} \tag{2.8}$$

The regular solution (where $\kappa > 0$) is obtained by taking the choice $\ell \to \gamma$ in the above parameter map. Using this choice in (2.8) and the well-known nonrelativistic energy spectrum, $E_n = -Z^2/2(\ell + n + 1)^2$, we obtain the following relativistic spectrum

$$\varepsilon_n = \pm\left[1 + \left(\frac{ƛZ}{n+\gamma+1}\right)^2\right]^{-1/2} \quad ; \qquad n = 0,1,2,... \tag{2.9}$$

The regular energy eigenfunction solution for the upper spinor component is obtained from the nonrelativistic wave function using the same parameter map (with $\ell \to \gamma$) as

$$\phi_n^+(r) = a_n \sqrt{\frac{\gamma/\kappa + \varepsilon_n}{\gamma/\kappa}} (\lambda_n r)^{\gamma+1} e^{-\lambda_n r/2} L_n^{2\gamma+1}(\lambda_n r) \tag{2.10}$$

where $\lambda_n = -2Z|\varepsilon_n|/(n+\gamma+1)$, $L_n^\nu(x)$ is the generalized Laguerre polynomial[18] and $a_n$ is the normalization constant $\sqrt{\lambda_n \Gamma(n+1)/2\Gamma(n+2\gamma+2)}$. The irregular solution (where $\kappa < 0$) is obtained by adopting the alternative choice $\ell \to -\gamma - 1$ in the parameter map (2.8). This gives the following energy spectrum and upper spinor component:

$$\bar{\varepsilon}_n = \pm\left[1 + \left(\frac{ƛZ}{n-\gamma}\right)^2\right]^{-1/2} \quad ; \quad n = 0,1,2,... \tag{2.11}$$

$$\bar{\phi}_n^+(r) = \bar{a}_n \sqrt{\frac{\gamma/\kappa + \bar{\varepsilon}_n}{\gamma/\kappa}} (\bar{\lambda}_n r)^{-\gamma} e^{-\bar{\lambda}_n r/2} L_n^{-2\gamma-1}(\bar{\lambda}_n r) \tag{2.12}$$

where $\bar{\lambda}_n = -2Z|\bar{\varepsilon}_n|/(n-\gamma)$ and $\bar{a}_n = \sqrt{\bar{\lambda}_n \Gamma(n+1)/2\Gamma(n-2\gamma)}$. On the other hand, the wave equation for the lower spinor component is obtained from Eq. (2.4) as

$$\left[-\frac{d^2}{dr^2} + \frac{\gamma(\gamma-1)}{r^2} + 2\frac{Z\varepsilon}{r} - \frac{\varepsilon^2 - 1}{ƛ^2}\right]\phi^-(r) = 0 \tag{2.13}$$

Comparing this with the Schrödinger-Coulomb wave equation (2.7) gives the following parameter map

$$\begin{aligned} &\ell \to \gamma - 1 \quad \text{or} \quad \ell \to -\gamma \\ &Z \to Z\varepsilon \\ &E \to (\varepsilon^2 - 1)/2ƛ^2 \end{aligned} \tag{2.14}$$

which results in the following regular (where, $\kappa > 0$ and $\ell \to \gamma - 1$) and irregular (where, $\kappa < 0$ and $\ell \to -\gamma$) lower spinor wave functions with energy eigenvalues $\varepsilon_{n-1}$ and $\bar{\varepsilon}_{n+1}$, respectively:

$$\phi_n^-(r) = a_{n-1}\sqrt{n(n+2\gamma)}\sqrt{\frac{\gamma/\kappa - \varepsilon_{n-1}}{\gamma/\kappa}}(\lambda_{n-1}r)^{\gamma}\, e^{-\lambda_{n-1}r/2} L_n^{2\gamma-1}(\lambda_{n-1}r)$$
$$\bar{\phi}_n^-(r) = \frac{-\bar{a}_{n+1}}{\sqrt{(n+1)(n-2\gamma+1)}}\sqrt{\frac{\gamma/\kappa - \bar{\varepsilon}_{n+1}}{\gamma/\kappa}}(\bar{\lambda}_{n+1}r)^{-\gamma+1}\, e^{-\bar{\lambda}_{n+1}r/2} L_n^{-2\gamma+1}(\bar{\lambda}_{n+1}r) \quad (2.15)$$

for $n = 0,1,2,..$. The lowest energy states with $\varepsilon_{-1} = -\bar{\varepsilon}_0 = -\gamma/\kappa$ are associated with the following two spinor wavefunctions, respectively

$$\Phi = \begin{pmatrix} 0 \\ \phi_0^- \end{pmatrix} = \sqrt{-2Z/\kappa\,\Gamma(2\gamma)}\left(-2Zr/\kappa\right)^{\gamma} e^{Zr/\kappa}\begin{pmatrix} 0 \\ 1 \end{pmatrix}$$
$$\bar{\Phi} = \begin{pmatrix} \bar{\phi}_0^+ \\ 0 \end{pmatrix} = \sqrt{2Z/\kappa\,\Gamma(-2\gamma)}\left(2Zr/\kappa\right)^{-\gamma} e^{-Zr/\kappa}\begin{pmatrix} 1 \\ 0 \end{pmatrix}$$

The second illustrative example is chosen to be the Dirac-Rosen-Morse problem where $V(r) = B\tanh(\lambda r)$, $B$ and $\lambda$ are real potential parameters. The potential constraint gives the odd potential component $W(r) = (B/\xi)\tanh(\lambda r) - \kappa/r = \pm(ƛB/S)\tanh(\lambda r) - \kappa/r$, where $S = \sin(ƛ\,\eta)$. Eq. (2.4), for the upper spinor component reads

$$\left[-\frac{d^2}{dr^2} - \frac{ƛB}{T}\left(\frac{ƛB}{T} + \lambda\right)\frac{1}{\cosh^2(\lambda r)} + 2\varepsilon B\tanh(\lambda r) + \left(\frac{ƛB}{T}\right)^2 - \frac{\varepsilon^2 - 1}{ƛ^2}\right]\phi^+(r) = 0 \quad (2.16)$$

where $T \equiv S/C = \tan(ƛ\eta)$. Comparing this with the nonrelativistic S-wave Rosen-Morse equation[19]

$$\left[-\frac{d^2}{dr^2} - A(A+\lambda)\frac{1}{\cosh^2(\lambda r)} + 2B\tanh(\lambda r) + A^2 - 2E\right]\psi(r) = 0 \quad (2.17)$$

gives the following parameter map:

$$\begin{aligned} &A \to ƛB/T \quad \text{or} \quad A \to -ƛB/T - \lambda \\ &B \to B\varepsilon \\ &E \to (\varepsilon^2 - 1)/2ƛ^2 \quad \text{or} \quad E \to (\varepsilon^2-1)/2ƛ^2 - \lambda(ƛB/T + \lambda/2) \end{aligned} \quad (2.18)$$

Using this parameter map for the regular solution (where $A \to ƛB/T$) and the known nonrelativistic energy spectrum,[19] we obtain the relativistic bound states energy spectrum as[20]

$$\varepsilon_n = \pm\left[\frac{1 + \left(ƛ^2 B/T\right)^2 - ƛ^2\lambda^2\left(ƛB/\lambda T - n\right)^2}{1 + \left(ƛB/\lambda\right)^2\left(ƛB/\lambda T - n\right)^{-2}}\right]^{1/2} \quad (2.19)$$

where $n = 0,1,2,...,n_{\max}$ and $n_{\max}$ is the largest integer satisfying

$$\left|n_{\max} - \frac{ƛB}{\lambda T}\right| < \frac{1}{ƛ\lambda}\sqrt{1 + ƛ^2\left(ƛB/T\right)^2}$$

Moreover, the same parameter map (with $A \to ƛB/T$) transforms the nonrelativistic wavefunction into the following regular solution for the upper spinor

$$\phi_n^+(r) = a_n\sqrt{(C+\varepsilon_n)/C}\,(1-z)^{(\mu_n+\nu_n)/2}(1+z)^{(\mu_n-\nu_n)/2} P_n^{(\mu_n+\nu_n,\mu_n-\nu_n)}(z) \quad (2.20)$$

where $P_n^{(\alpha,\beta)}(z)$ is the Jacobi polynomial,[18] and

$$
\begin{aligned}
&z = \tanh(\lambda r)\\
&\mu_n = ƛB/\lambda T - n\\
&\nu_n = B\varepsilon_n/\lambda^2\mu_n \qquad\qquad (2.21)\\
&a_n = \left[\frac{\lambda}{2}\frac{n+\mu_n+1/2}{2^{2\mu_n}}\frac{\Gamma(n+1)\Gamma(n+2\mu_n+1)}{\Gamma(n+\mu_n+\nu_n+1)\Gamma(n+\mu_n-\nu_n+1)}\right]^{1/2}
\end{aligned}
$$

Now, the second order differential equation for the lower spinor component reads

$$
\left[-\frac{d^2}{dr^2}-\frac{ƛB}{T}\left(\frac{ƛB}{T}-\lambda\right)\frac{1}{\cosh^2(\lambda r)}+2\varepsilon B\tanh(\lambda r)+\left(\frac{ƛB}{T}\right)^2-\frac{\varepsilon^2-1}{ƛ^2}\right]\phi^-(r)=0 \quad (2.22)
$$

Comparing it with the nonrelativistic S-wave Rosen-Morse equation (2.17) gives the following parameter map:

$$
\begin{aligned}
&A \to ƛB/T-\lambda \quad \text{or} \quad A\to -ƛB/T\\
&B \to B\varepsilon \qquad\qquad (2.23)\\
&E \to (\varepsilon^2-1)/2ƛ^2+\lambda(ƛB/T-\lambda/2) \quad \text{or} \quad E\to(\varepsilon^2-1)/2ƛ^2
\end{aligned}
$$

The choice of map with $A \to \alpha B/T - \lambda$ results in the following regular solution for the lower spinor component whose energy eigenvalue is $\varepsilon_{n+1}$

$$
\begin{aligned}
\phi_n^-(r) = a_{n+1}&\left[\frac{n+\mu_{n+1}+1/2}{n+\mu_{n+1}+3/2}\frac{(n+\mu_{n+1}-\nu_{n+1}+1)(n+\mu_{n+1}+\nu_{n+1}+1)}{(n+1)(n+2\mu_{n+1}+1)}\right]^{1/2} \qquad (2.24)\\
&\times\sqrt{(C-\varepsilon_{n+1})/C}\,(1-z)^{(\mu_{n+1}-\nu_{n+1})/2}(1+z)^{(\mu_{n+1}+\nu_{n+1})/2}P_n^{(\mu_{n+1}-\nu_{n+1},\,\mu_{n+1}+\nu_{n+1})}(z)
\end{aligned}
$$

where $n = 0,1,2,...,n_{\max}-1$. The irregular solution is obtained in the same manner using the alternative choice of parameter map. That is, by using the map in which $A \to -ƛB/T - \lambda$ (respectively, $A \to -ƛB/T$) for the upper (respectively, lower) spinor component.

A comprehensive list of all known (up to date) relativistic problems that are solvable by this approach is given in Table 1. For each problem, the Table lists the relativistic odd and even potential components, the transformation parameter, the bound states energy spectrum for the regular solution, and any relations among the parameters. To generate the parameter maps, Table 2 lists the 2nd order differential equations for the upper spinor components and their corresponding nonrelativistic Schrödinger equations. Table 3 lists the regular solution for upper spinor radial wavefunctions. The lower spinor components can be easily obtained starting from Eq. (2.4), with the lower sign, and following the same procedure as shown in the above two examples. The irregular solutions could, as well, be generated in an analogous manner as depicted above in the given examples.

### 3. Potential-Class Invariant Maps

The dynamical symmetry of a physical problem could be exploited by studying the representations of its spectrum generating algebra. Realization of the generators of such algebra by differential operators facilitates study of the properties of the wave equation and its solutions. Potentials with symmetry that is associated with such algebra are grouped into classes. All problems in a given class could be mapped into one another by point canonical transformation (PCT). An example is the so(2,1) Lie algebra which is the potential algebra for many three parameter problems in nonrelativistic quantum mechanics.[21] Potentials associated with this algebra include, but not limited to, the oscillator, Coulomb, and S-wave Morse. The wavefunction solutions of this class of problems are all written in terms of the confluent hypergeometric function. All of these potentials can be mapped into the oscillator potential by PCTs. For this reason, this class is sometimes referred to as the "oscillator class". A graded extension of so(2,1) Lie algebra was defined in Ref. 11. Its realization by two-dimensional matrices of differential operators acting in the spinor space was given therein. The linear span of this superalgebra gives the canonical form of the radial Dirac Hamiltonian, which carries a representation of this supersymmetry. It turned out that this graded algebra is the supersymmetry algebra for the class of relativistic potentials that includes the Dirac-Oscillator, Dirac-Coulomb and Dirac-Morse potentials. It is, in fact, the relativistic extension of the oscillator class. A new point canonical transformation (XPCT), which is compatible with the relativistic wave equation, was formulated. It preserves the canonical form of the Dirac equation. Therefore, just like PCT in the nonrelativistic theory, XPCT maps the solution of a reference problem into other exact solutions of the Dirac equation that belong to the same class as the reference problem. In fact, XPCT maps the relativistic problems mentioned above into the Dirac-Oscillator problem. Therefore, if one starts with a relativistic problem whose solution is well established, applying XPCTs that preserve the structure of the wave equation will result in new solutions of the equation. Thus the reference problem acts like a seed for generating exact solutions. This scheme is suitable in the search for solutions of a given wave equation (e.g., the Dirac equation) by making an exhaustive study of all XPCTs that maintain shape invariance of the given equation. For clarity of the following presentation, we choose a notation where the configuration space coordinates for the reference problem is $\rho$ while that for the "target problem" is $r$. Moreover, common parameters and variables are distinguished with the caret symbol for those belonging to the reference problem. XPCT is the transformation

$$\begin{aligned} & r = q(\rho) \\ & \phi^{\pm}(r) = g^{\pm}(\rho)\hat{\phi}^{\pm}(\rho) \end{aligned} \tag{3.1}$$

that leaves the relativistic wave equation (2.2) form-invariant. $q(\rho)$ and $g^{\pm}(\rho)$ are real differentiable functions. Hermiticity and Schrödinger-like requirements give $g^{\pm}(\rho) = \sqrt{q'\left[(\hat{C}\pm\hat{\varepsilon})/(C\pm\varepsilon)\right]}$. Moreover, covariance of the wave equation results in the following condition

$$\mp\xi + CU(r) = \frac{1}{q'}\left[\mp\hat{\xi} + \hat{C}\hat{U}(\rho) \mp \frac{1}{2}\frac{q''}{q'}\right] \tag{3.2}$$

where $q' \equiv dq/d\rho$, $U(r) = W(r) + \kappa/r$ and similarly $\hat{U}(\rho) = \hat{W}(\rho) + \hat{\kappa}/\rho$. Eq. (3.2) is an equality relation modulo a constant due to the derivatives. It gives the spin-orbit coupling $\kappa$ of the target problem in terms of the reference problem parameters. It also results in an expression for the odd component of the relativistic potential, $W(r)$, in terms of the reference problem quantities and the transformation function $q$. Equating the determinants of the two wave operators (equivalently, the corresponding 2nd order differential operators) and using Eq. (3.2) results in the following relation

$$C^2U^2 + 2\xi\varepsilon U - \frac{\varepsilon^2 - 1}{\lambda^2} =$$

$$\frac{1}{(q')^2}\left[\hat{C}^2\hat{U}^2 + 2\hat{\xi}\hat{\varepsilon}\hat{U} - \frac{\hat{\varepsilon}^2 - 1}{\lambda^2} + \frac{1}{4}\left(\frac{q''}{q'}\right)^2 + \frac{q''}{q'}\left(\hat{\xi} \mp \hat{C}\hat{U}\right)\right] \tag{3.3}$$

which gives the energy $\varepsilon$, thus completing the map among all parameters of the two problems. Using this parameter map and the known solution of the reference problem (energy spectrum and spinor wave functions), we obtain the solution of the target problem. As an example, we will give next a brief account of the Dirac-Oscillator class, its superalgebra and the associated XPCT map.

so(2,1) Lie algebra is a three dimensional algebra whose basis elements satisfy the commutation relations $[L_3, L_\pm] = \pm L_\pm$, $[L_+, L_-] = -L_3$. We define a graded extension of this algebra as the four dimensional superalgebra with two odd elements $L_\pm$ and two even elements $L_0$, $L_3$ satisfying the commutation/anticommutation relations $[L_3, L_\pm]$ = $\pm L_\pm$, $\{L_+, L_-\} = L_0$, and $[L_0, L_3] = [L_0, L_\pm] = 0$. $L_0$ forms the center of this superalgebra since it commutes with all of its elements. A realization of the generators of this superalgebra by 2×2 matrices of differential operators acting in the two-component spinor space was constructed in Ref. 11. The odd operators, which are the raising and lowering operators in this two dimensional spinor space, are first order in the derivatives, whereas the even operators are zero and second order. Requiring that the linear span of this graded algebra be hermitian and first order differential gives the following 2×2 radial differential operator

$$\begin{pmatrix} 1 & \lambda\left[U(r) - \frac{d}{dr}\right] \\ \lambda\left[U(r) + \frac{d}{dr}\right] & -1 \end{pmatrix} \tag{3.4}$$

where $U(r)$ is a real function. This is identical to the *canonical* form of the Dirac Hamiltonian in Eq. (2.2) with $\eta = 0$ [equivalently, Eq. (2.1) with $V(r) = 0$] and $U(r) = W(r) + \kappa/r$. This canonical form of the Dirac Hamiltonian is compatible with the Dirac-Oscillator reference problem defined (as shown in Table 1) with $\eta = 0$ and $W(r) = \omega^2 r$, where $\omega$ is the oscillator frequency. It is to be noted that an analogous feature also appears in the nonrelativistic theory. The linear span of so(2,1) Lie algebra results in a canonical form for the Schrödinger operator which is compatible with the nonrelativistic oscillator problem (See, for example, the Appendix in Ref. 10). Moreover, since $L_0$ is the center of the superalgebra (i.e. a constant scalar) then the two-component spinor must belong to the space of its eigenvectors. That is $L_0\Phi \propto \Phi$. Therefore, using the differential matrix realization for this operator,[11] we can write

$$\begin{pmatrix} -\frac{d^2}{dr^2} + U^2 - U' & 0 \\ 0 & -\frac{d^2}{dr^2} + U^2 + U' \end{pmatrix}\begin{pmatrix} \phi^+ \\ \phi^- \end{pmatrix} = \frac{\varepsilon^2 - 1}{\lambda^2}\begin{pmatrix} \phi^+ \\ \phi^- \end{pmatrix} \tag{3.5}$$

Now, the solution of the Dirac-Oscillator problem could also be obtained following the same procedure used in the examples of Sec. 2. The regular and irregular solution has the following spinor wavefunctions, respectively

$$\Phi_n = \begin{pmatrix} \phi_n^+ \\ \phi_n^- \end{pmatrix} = (\omega r)^\kappa e^{-\omega^2 r^2/2} \begin{pmatrix} a_n^\kappa \sqrt{1+\varepsilon_n}(\omega r) L_n^{\kappa+1/2}(\omega^2 r^2) \\ a_n^{\kappa-1} \sqrt{1-\varepsilon_n} L_n^{\kappa-1/2}(\omega^2 r^2) \end{pmatrix} \tag{3.6}$$

$$\bar{\Phi}_n = \begin{pmatrix} \bar{\phi}_{n+1}^+ \\ \bar{\phi}_n^- \end{pmatrix} = (\omega r)^{-\kappa} e^{-\omega^2 r^2/2} \begin{pmatrix} a_{n+1}^{-\kappa-1} \sqrt{1+\bar{\varepsilon}_{n+1}} L_{n+1}^{-\kappa-1/2}(\omega^2 r^2) \\ -a_n^{-\kappa} \sqrt{1-\bar{\varepsilon}_{n+1}}(\omega r) L_n^{-\kappa+1/2}(\omega^2 r^2) \end{pmatrix} \tag{3.7}$$

where $n = 0,1,2,\ldots$ and $a_n^\kappa = \sqrt{\omega \Gamma(n+1)/\Gamma(n+\kappa+3/2)}$. The respective energy spectra are as follows:

$$\begin{aligned} \varepsilon_n &= \pm\sqrt{1+4\lambda^2\omega^2(n+\kappa+1/2)} \\ \bar{\varepsilon}_n &= \pm\sqrt{1+4\lambda^2\omega^2 n} \end{aligned} \tag{3.8}$$

The lowest energy state, where $\bar{\varepsilon}_0 = 1$, is associated with the following irregular spinor wavefunction:

$$\bar{\Phi} = \begin{pmatrix} \bar{\phi}_0^+ \\ 0 \end{pmatrix} = \sqrt{2\omega/\Gamma(-\kappa+1/2)}(\omega r)^{-\kappa} e^{-\omega^2 r^2/2} \begin{pmatrix} 1 \\ 0 \end{pmatrix}$$

To obtain the solution of all other problems in the class, one merely needs to make proper choices of the XPCT function, $q(\rho)$, that preserve the functional form of the Dirac equation. That is, the choice of $q(\rho)$ should be compatible with the relations (3.2) and (3.3). The implementation steps of this scheme are illustrated in Ref. 11. Taking $q(\rho) = \rho^2$ solves the Dirac-Coulomb problem, while the choice $q(\rho) = -(2/\tau)\ln\rho$ results in a solution of the S-wave Dirac-Morse problem with $\tau$ being the range parameter. Choosing $q(\rho) = \rho^{\nu+1}$, where $\nu \neq 0, \pm 1$, gives the solution of the relativistic problem with power-law potential at rest mass energy ($\varepsilon$ = 1).[10] As an illustrative example, we will generate the solution of the S-wave Dirac-Morse problem by applying the XPCT on the Dirac-Oscillator. Substituting the XPCT function $q(\rho) = -(2/\tau)\ln\rho$ in Eq. (3.2) gives $W(r) = -\left(\tau\omega^2/2C\right)e^{-\tau r} - \kappa/r$. On the other hand, Eq. (3.3) give:

$$\begin{aligned} \mp\frac{T\varepsilon}{\lambda} &= -\tau\hat{\mu} + \frac{\tau}{2}\left(\hat{\kappa} \pm \tfrac{1}{2}\right) \\ \frac{\varepsilon^2-1}{\lambda^2} &= -\frac{\tau^2}{4}\left(\hat{\kappa} \pm \tfrac{1}{2}\right)^2 \end{aligned} \tag{3.9}$$

where the top/bottom sign refers to the upper/lower spinor component and the parameter $\hat{\mu}$ is defined as $\hat{\mu} \equiv (\hat{\varepsilon}^2 - 1)/4\lambda^2\omega^2$ which may depend on $\hat{\kappa}$. The first equation in (3.9) is solved for $\hat{\kappa} \pm \tfrac{1}{2}$ which when substituted in the second gives a quadratic equation in $\varepsilon$. The result for the regular solution is

$$\begin{aligned} \varepsilon_n &= \left\{-\lambda\tau T(n+\tfrac{1}{2}) \pm \sqrt{1+T^2-[\lambda\tau T(n+\tfrac{1}{2})]^2}\right\}\Big/(1+T^2) \\ \varepsilon_n^+ &= \varepsilon_n \quad \text{and} \quad \varepsilon_n^- = -\varepsilon_n \\ \hat{\kappa} &= \pm\left(2T\varepsilon_n/\lambda\tau + \tfrac{1}{2}\right) - 2n - 1 \end{aligned} \tag{3.10}$$

while, for the irregular solution it reads

$$\bar{\varepsilon}_n = \left[ \text{ƛ}\tau T n \pm \sqrt{1+T^2-(\text{ƛ}\tau T n)^2}\, \right] \Big/ (1+T^2)$$

$$\bar{\varepsilon}_n^+ = \bar{\varepsilon}_n \quad \text{and} \quad \bar{\varepsilon}_n^- = -\bar{\varepsilon}_{n+1} \tag{3.11}$$

$$\hat{\kappa} = \mp\left(2T\bar{\varepsilon}_n/\text{ƛ}\tau + ½\right) + 2n + 1 \mp 1$$

The spinor wavefunctions are obtained, up to normalization constant, by the XPCT map (3.1) as $\phi^{\pm}(r) = \sqrt{q'\left[(\hat{C}\pm\hat{\varepsilon})/(C\pm\varepsilon)\right]}\hat{\phi}^{\pm}(\rho)$. The results are as follows:

$$\Phi_n = \begin{pmatrix} \phi_n^+ \\ \phi_n^- \end{pmatrix} = \begin{pmatrix} a_n^+ \sqrt{(C+\varepsilon_n)/C}\, x^{\nu_n} e^{-x/2} L_n^{2\nu_n}(x) \\ -a_n^- \sqrt{(C-\varepsilon_n)/C}\, x^{\nu_n-1} e^{-x/2} L_n^{2(\nu_n-1)}(x) \end{pmatrix}; \quad n = 0,1,2,\ldots,n_{\max} \tag{3.12}$$

$$\bar{\Phi}_n = \begin{pmatrix} \bar{\phi}_{n+1}^+ \\ \bar{\phi}_n^- \end{pmatrix} = \begin{pmatrix} \bar{a}_{n+1}^+ \sqrt{(C+\bar{\varepsilon}_{n+1})/C}\, x^{\mu_n} e^{-x/2} L_{n+1}^{2\mu_n}(x) \\ -\bar{a}_n^- \sqrt{(C-\bar{\varepsilon}_{n+1})/C}\, x^{\mu_n} e^{-x/2} L_n^{2\mu_n}(x) \end{pmatrix}; \quad n = 0,1,2,\ldots,\bar{n}_{\max} \tag{3.13}$$

where $x = \omega^2 e^{-\tau r}$, $\nu_n = \frac{T\varepsilon_n}{\text{ƛ}\tau} - n$, $\mu_n = \frac{T\bar{\varepsilon}_{n+1}}{\text{ƛ}\tau} - n - 1 = \bar{\nu}_{n+1}$, $n_{\max} \le \sqrt{1+T^{-2}}\big/\text{ƛ}\tau - 1/2$ and $\bar{n}_{\max} \le \sqrt{1+T^{-2}}\big/\text{ƛ}\tau - 1$. The lowest energy state, where $\bar{\varepsilon}_0 = 1\big/\sqrt{1+T^2} = C$, is associated with the following irregular spinor wavefunction:

$$\bar{\Phi} = \begin{pmatrix} \bar{\phi}_0^+ \\ 0 \end{pmatrix} = \sqrt{2}\bar{a}_0^+ x^{S/\text{ƛ}\tau} e^{-x/2} \begin{pmatrix} 1 \\ 0 \end{pmatrix}$$

The spinor wavefunctions for all problems in the Dirac-Oscillator class are written in terms of the confluent hypergeometric function. Another class of exactly solvable relativistic potentials, whose spinor wave functions are written in terms of the hypergeometric function,[9] includes Dirac-Rosen-Morse I & II, Dirac-Eckart, Dirac-Pöschl-Teller, and Dirac-Scarf potentials. It also carries a representation of the graded so(2,1) superalgebra. A similar XPCT treatment could be carried out for this class where the reference problem is taken to be associated with any suitable one of these potentials, e.g. the Dirac-Rosen-Morse. Table 4 lists these two classes along with the XPCT function $q(\rho)$ for each potential.

## 4. Summary

In conclusion, there are two ways to obtain solutions of all relativistic problems in a given potential class using the approach presented in this work. If the solution of any one of the problems in the class is known then, as shown in Sec. 3, applying XPCT maps that preserve the structure of the wave equation on this solution will produce the rest. However, if no solution is known, then any one of the problems should be solved by correspondence with a well known exactly solved nonrelativistic problem as shown in Sec. 2. Thereafter, XPCT maps this solution into those of the remaining problems in the class.

## Acknowledgements

The author is grateful to Amjad A. Al-Haidari for valuable assistance in compiling and organizing the material used in support of this work.

## REFERENCES:

1. See, for example, G. A. Natanzon, *Teor. Mat. Fiz.* **38**, 219 (1979) [*Theor. Math. Phys.* **38**, 146 (1979)]; L. E. Gendenshtein, *Zh. Eksp. Teor. Fiz. Pis'ma Red.* **38**, 299 (1983) [*JETP Lett.* **38**, 356 (1983)]; F. Cooper, J. N. Ginocchi, and A. Khare, *Phys. Rev.* D **36**, 2438 (1987); R. Dutt, A. Khare, and U. P. Sukhatme, *Am. J. Phys.* **56**, 163 (1988); *ibid.* **59**, 723 (1991); G. Lévai, *J. Phys.* A **22**, 689 (1989); *ibid.* **27**, 3809 (1994)
2. M. F. Manning, *Phys. Rev.* **48**, 161 (1935); A. Bhattacharjie and E. C. G. Sudarshan, *Nuovo Cimento* **25**, 864 (1962); N. K. Pak and I. Sökmen, *Phys. Lett.* **103A**, 298 (1984); H. G. Goldstein, *Classical Mechanics* (Addison-Wesley, Reading-MA, 1986); R. Montemayer, *Phys. Rev.* A **36**, 1562 (1987); G. Junker, *J. Phys.* A **23**, L881 (1990)
3. See, for example, E. Witten, *Nucl. Phys.* B **185**, 513 (1981); F. Cooper and B. Freedman, *Ann. Phys.* (NY) **146**, 262 (1983); C. V. Sukumar, *J. Phys.* A **18**, 2917 (1985); A. Arai, *J. Math. Phys.* **30**, 1164 (1989); F. Cooper, A. Khare, and U. Sukhatme, *Phys. Rep.* **251**, 267 (1995)
4. See, for example, B. G. Wybourne, *Classical groups for physicists* (Wiley-Interscience, New York, 1974), and references therein; W. Miller Jr., *Lie theory and special functions* (Academic, New York, 1968); Y. Alhassid, F. Gürsey, and F. Iachello, *Phys. Rev. Lett.* **5**0, 873 (1983); *Ann. Phys.* (N.Y.) **148**, 346 (1983) ; *ibid.* **16**7, 181 (1986); Y. Alhassid, F. Iachello, and R. Levine, *Phys. Rev. Lett.* **54**, 1746 (1985); Y. Alhassid, F. Iachello, and J. Wu, *Phys. Rev. Lett.* **56**, 271 (1986); J. Wu and Y. Alhassid, *J. Math. Phys.* **31**, 557 (1990); M. J. Englefield and C. Quesne, *J. Phys.* A **24**, 3557 (1991)
5. A. de Souza-Dutra, *Phys. Rev.* A **4**7, R2435 (1993); N. Nag, R. Roychoudhury, and Y. P. Varshni, *Phys. Rev.* A **49**, 5098 (1994); R. Dutt, A. Khare, and Y. P. Varshni, *J. Phys.* A **28**, L107 (1995); C. Grosche, *J. Phys.* A, **28**, 5889 (1995); *ibid.* **29**, 365 (1996); G. Lévai and P. Roy, *Phys. Lett.* A **270**, 155 (1998); G. Junker and P. Roy, *Ann. Phys.* (N.Y.) **26**4, 117 (1999); R. Roychoudhury, P. Roy, M. Zonjil, and G. Lévai, *J. Math. Phys.* **42**, 1996 (2001)
6. A. V. Turbiner, *Commun. Math. Phys.* **118**, 467 (1988); M. A. Shifman, *Int. J. Mod. Phys.* A **4**, 2897 (1989); R. Adhikari, R. Dutt, and Y. Varshni, *Phys. Lett.* A **141**, 1 (1989); *J. Math. Phys.* **32**, 447 (1991); R. K. Roychoudhury, Y. P. Varshni, and M. Sengupta, *Phys. Rev.* A **42**, 184 (1990); L. D. Salem and R. Montemayor, *Phys. Rev.* A **43**, 1169 (1991); M. W. Lucht and P. D. Jarvis, *Phys. Rev.* A **47**, 817 (1993); A. G. Ushveridze, *Quasi-exactly Solvable Models in Quantum Mechanics* (IOP, Bristol, 1994)
7. M. Moshinsky and A. Szczepaniak, *J. Phys.* A **22**, L817 (1989); J. Bentez, R. P. Martinez-y-Romero, H. N. Nunez-Yepez, and A. L. Salas-Brito, *Phys. Rev. Lett.* **64**, 1643 (1990); O. L. de Lange, *J. Phys.* A **24**, 667 (1991); V. M. Villalba, *Phys. Rev.* A **49**, 586 (1994); P. Rozmej and R. Arvieu, *J. Phys.* A **32**, 5367 (1999); R. Szmytkowski and M. Gruchowski, *J. Phys.* A **34**, 4991 (2001)
8. A. D. Alhaidari, *Phys. Rev. Lett.* **87**, 210405 (2001); **88**, 189901 (2002)
9. A. D. Alhaidari, *J. Phys.* A **34**, 9827 (2001); **35**, 6207 (2002)
10. A. D. Alhaidari, *Int. J. Mod. Phys.* A **17**, 4551 (2002)
11. A. D. Alhaidari, *Phys. Rev.* A **65**, 042109 (2002); **66**, 019902 (2002)
12. A. N. Vaidya and R. de L. Rodrigues, *Phys. Rev. Lett.* **89**, 068901 (2002)
13. A. S. de Castro, *J. Phys.* A, **35**, 6203 (2002)
14. A. N. Vaidya and R. de L. Rodrigues, *J. Phys.* A **35**, 6205 (2002)

15. G. Jian-You, F. Xiang Zheng, and Xu Fu-Xin, *Phys. Rev.* A **66**, 062105 (2002)
16. J. D. Bjorken and S. D. Drell, *Relativistic Quantum Mechanics* (McGraw Hill, New York, 1965)
17. A. D. Alhaidari, to be published; e-print hep-th/0207018
18. W. Magnus, F. Oberhettinger, and R. P. Soni, *Formulas and Theorems for the Special Functions of Mathematical Physics*, 3rd edition (Springer-Verlag, New York, 1966)
19. R. De, R. Dutt, and U. Sukhatme, *J. Phys.* A **25**, L843 (1992)
20. The expressions for the relativistic energy spectrum of the Dirac-Rosen-Morse I and Dirac-Eckart problems in Ref. 9 contain a typo. The correction is as shown here and in Table 1.
21. B. G. Wybourne, *Classical groups for physicists* (Wiley-Interscience, New York, 1974), and references therein; W. Miller Jr., *Lie theory and special functions* (Academic, New York, 1968); Y. Alhassid, F. Gürsey, and F. Iachello, *Phys. Rev. Lett.* **5**0, 873 (1983); *Ann. Phys.* (N.Y.) **148**, 346 (1983); *ibid.* **16**7, 181 (1986); Y. Alhassid, F. Iachello, and J. Wu, *Phys. Rev. Lett.* **56**, 271 (1986); J. Wu and Y. Alhassid, *J. Math. Phys.* **31**, 557 (1990); M. J. Englefield and C. Quesne, *J. Phys.* A **24**, 3557 (1991)
22. S. Flügge, *Practical Quantum Mechanics* (Springer-Verlag, Berlin, 1974) p. 165

## TABLE CAPTIONS

**Table 1:** A comprehensive list of all known (up to date) relativistic problems, which are exactly solvable by the approach. For each problem, the Table gives the relativistic odd and even potential components $W(r)$ and $V(r)$, the transformation parameter $\eta$, the bound states energy spectrum $\varepsilon_n$ for the regular solution, and any relations among the parameters. For the nonrelativistic Woods-Saxon problem, the energy spectrum is obtained by the boundary constraint $f\left(E,a,R,V_0\right) = (n - ½)\pi$,[22] where

$$f\left(E,a,R,V_0\right) = \nu R/a - \tan^{-1}(\nu/\mu) + \arg\Gamma(2i\nu) - 2\arg\Gamma(\mu + i\nu)$$

$$\mu^2 = -2a^2 E\,, \qquad \nu^2 = 2a^2 V_0 - \mu^2$$

and $n = 0, \pm1, \pm2, \ldots$.

**Table 2:** A list of the second order differential equations for the upper spinor components and their corresponding nonrelativistic Schrödinger equations. These are used to generate the parameter maps shown for the regular solutions.

**Table 3:** A list of the regular solutions for upper spinor radial wavefunctions. The lower spinor components can be obtained starting from Eq. (2.4) and following the same procedure as shown in the two examples of Sec. 2.

**Table 4:** A list of relativistic potentials in the Dirac-Oscillator and Dirac-Rosen-Morse classes along with the XPCT function, $q(\rho)$, for each potential in its respective class.

**Table 1**

| | $V(r)$ | $W(r)$ | $\pm\eta$ | $\varepsilon_n$ | Parameters Relation |
|---|---|---|---|---|---|
| **Dirac-Coulomb** | $Z/r$ | $0$ | $\sin^{-1}(\bar{\lambda}Z/\kappa)$ | $\left\{1+\left[\bar{\lambda}Z/(\gamma+n+1)\right]^2\right\}^{-1/2}$ | $\kappa^2=\gamma^2+\bar{\lambda}^2Z^2$ |
| **Dirac-Oscillator** | $0$ | $\omega^2 r$ | $0$ | $\sqrt{1+2\bar{\lambda}^2\omega^2(2n+\ell+\kappa+1)}$ | $\kappa=\ell,-\ell-1$ |
| **Dirac-Morse** | $-Be^{-\lambda r}$ | $-\tau e^{-\lambda r}-\kappa/r$ | $\sin^{-1}(\bar{\lambda}B/\tau)$ | $A\tau^{-2}\left[\bar{\lambda}^2\lambda Bn+\sqrt{\tau^2-(\bar{\lambda}\lambda An)^2}\right]$ | $\tau^2=A^2+\bar{\lambda}^2B^2$ |
| **Dirac-Rosen-Morse I** | $B\tanh(\lambda r)$ | $\tau\tanh(\lambda r)-\kappa/r$ | $\sin^{-1}(\bar{\lambda}B/\tau)$ | $\left[\dfrac{1+\bar{\lambda}^2A^2-\bar{\lambda}^2\lambda^2\left(A/\lambda-n\right)^2}{1+\left(\bar{\lambda}B/\lambda\right)^2\left(A/\lambda-n\right)^{-2}}\right]^{1/2}$ | $\tau^2=A^2+\bar{\lambda}^2B^2$ |
| **Dirac-Eckart** | $B\coth(\lambda r)$ | $\tau\coth(\lambda r)-\kappa/r$ | $\sin^{-1}(\bar{\lambda}B/\tau)$ | $\left[\dfrac{1+\bar{\lambda}^2A^2-\bar{\lambda}^2\lambda^2\left(A/\lambda-n\right)^2}{1+\left(\bar{\lambda}B/\lambda\right)^2\left(A/\lambda-n\right)^{-2}}\right]^{1/2}$ | $\tau^2=A^2+\bar{\lambda}^2B^2$ |
| **Dirac-Rosen-Morse II** | $0$ | $A\coth(\lambda r)-B\,\mathrm{csch}(\lambda r)-\kappa/r$ | $0$ | $\sqrt{1+\bar{\lambda}^2A^2-\bar{\lambda}^2\lambda^2\left(A/\lambda-n\right)^2}$ | ——— |
| **Dirac-Scarf** | $0$ | $A\tanh(\lambda r)+B\,\mathrm{sech}(\lambda r)-\kappa/r$ | $0$ | $\sqrt{1+\bar{\lambda}^2A^2-\bar{\lambda}^2\lambda^2\left(A/\lambda-n\right)^2}$ | ——— |
| **Dirac-Pöschl-Teller** | $0$ | $-A\tanh(\lambda r)-B\coth(\lambda r)-\kappa/r$ | $0$ | $\sqrt{1+\bar{\lambda}^2(A+B)^2-\bar{\lambda}^2\lambda^2\left[(A+B)/\lambda+2n\right]^2}$ | ——— |
| **Dirac-Woods-Saxon** | $\dfrac{-B}{1+e^{\lambda(r-R)}}$ | $\dfrac{\tau}{1+e^{\lambda(r-R)}}-\kappa/r$ | $-\sin^{-1}(\bar{\lambda}B/\tau)$ | $f\left((\varepsilon^2-1)/2\bar{\lambda}^2,\lambda^{-1},R,\varepsilon B-\lambda^2/2\right)=(n+½)\pi$ | $\tau^2=\lambda^2+\bar{\lambda}^2B^2$ |

**Table 2**

| | Relativistic 2nd order differential equation | Nonrelativistic Schrödinger equation | Parameters Map |
|---|---|---|---|
| **Dirac-Coulomb** | $\left[-\frac{d^2}{dr^2}+\frac{\gamma(\gamma+1)}{r^2}+2\frac{Z\varepsilon}{r}-\frac{\varepsilon^2-1}{\bar{\lambda}^2}\right]\phi^+=0$ | $\left[-\frac{d^2}{dr^2}+\frac{\ell(\ell+1)}{r^2}+2\frac{Z}{r}-2E\right]\psi=0$ | $\ell\to\gamma$<br>$Z\to Z\varepsilon$<br>$E\to(\varepsilon^2-1)/2\bar{\lambda}^2$ |
| **Dirac-Oscillator** | $\left[-\frac{d^2}{dr^2}+\frac{\kappa(\kappa+1)}{r^2}+\omega^4r^2+\omega^2(2\kappa-1)-\frac{\varepsilon^2-1}{\bar{\lambda}^2}\right]\phi^+=0$ | $\left[-\frac{d^2}{dr^2}+\frac{\ell(\ell+1)}{r^2}+\omega^4r^2-2E\right]\psi=0$ | $\ell\to\kappa,\omega\to\omega$<br>$E\to(\varepsilon^2-1)/2\bar{\lambda}^2$<br>$+\omega^2(½-\kappa)$ |
| **Dirac-Morse** | $\left[-\frac{d^2}{dr^2}+A^2e^{-2\lambda r}-A\left(2\varepsilon B/A+\lambda\right)e^{-\lambda r}-\frac{\varepsilon^2-1}{\bar{\lambda}^2}\right]\phi^+=0$ | $\left[-\frac{d^2}{dr^2}+A^2e^{-2\lambda r}-A\left(2B+\lambda\right)e^{-\lambda r}-2E\right]\psi=0$ | $A\to A,B\to\varepsilon B/A$<br>$E\to(\varepsilon^2-1)/2\bar{\lambda}^2$ |
| **Dirac-Rosen-Morse I** | $\left[-\frac{d^2}{dr^2}-\frac{A(A+\lambda)}{\cosh(\lambda r)^2}+2\varepsilon B\tanh(\lambda r)+A^2-\frac{\varepsilon^2-1}{\bar{\lambda}^2}\right]\phi^+=0$ | $\left[-\frac{d^2}{dr^2}-\frac{A(A+\lambda)}{\cosh(\lambda r)^2}+2B\tanh(\lambda r)+A^2-2E\right]\psi=0$ | $A\to A,B\to\varepsilon B$<br>$E\to(\varepsilon^2-1)/2\bar{\lambda}^2$ |
| **Dirac-Eckart** | $\left[-\frac{d^2}{dr^2}+\frac{A(A+\lambda)}{\sinh(\lambda r)^2}+2\varepsilon B\coth(\lambda r)+A^2-\frac{\varepsilon^2-1}{\bar{\lambda}^2}\right]\phi^+=0$ | $\left[-\frac{d^2}{dr^2}+\frac{A(A+\lambda)}{\sinh(\lambda r)^2}+2B\coth(\lambda r)+A^2-2E\right]\psi=0$ | $A\to A,B\to\varepsilon B$<br>$E\to(\varepsilon^2-1)/2\bar{\lambda}^2$ |
| **Dirac-Rosen-Morse II** | $\left[-\frac{d^2}{dr^2}+\frac{A^2+B^2+\lambda A}{\sinh(\lambda r)^2}-B(2A+\lambda)\frac{\cosh(\lambda r)}{\sinh(\lambda r)^2}+A^2-\frac{\varepsilon^2-1}{\bar{\lambda}^2}\right]\phi^+=0$ | $\left[-\frac{d^2}{dr^2}+\frac{A^2+B^2+\lambda A}{\sinh(\lambda r)^2}-B(2A+\lambda)\frac{\cosh(\lambda r)}{\sinh(\lambda r)^2}+A^2-2E\right]\psi=0$ | $A\to A,B\to B$<br>$E\to(\varepsilon^2-1)/2\bar{\lambda}^2$ |
| **Dirac-Scarf** | $\left[-\frac{d^2}{dr^2}-\frac{A^2-B^2+\lambda A}{\cosh(\lambda r)^2}+B(2A+\lambda)\frac{\sinh(\lambda r)}{\cosh(\lambda r)^2}+A^2-\frac{\varepsilon^2-1}{\bar{\lambda}^2}\right]\phi^+=0$ | $\left[-\frac{d^2}{dr^2}-\frac{A^2-B^2+\lambda A}{\cosh(\lambda r)^2}+B(2A+\lambda)\frac{\sinh(\lambda r)}{\cosh(\lambda r)^2}+A^2-2E\right]\psi=0$ | $A\to A,B\to B$<br>$E\to(\varepsilon^2-1)/2\bar{\lambda}^2$ |
| **Dirac-Pöschl-Teller** | $\left[-\frac{d^2}{dr^2}+\frac{B(B-\lambda)}{\sinh(\lambda r)^2}-\frac{A(A-\lambda)}{\cosh(\lambda r)^2}+(A+B)^2-\frac{\varepsilon^2-1}{\bar{\lambda}^2}\right]\phi^+=0$ | $\left[-\frac{d^2}{dr^2}+\frac{B(B-\lambda)}{\sinh(\lambda r)^2}-\frac{A(A-\lambda)}{\cosh(\lambda r)^2}+(A+B)^2-2E\right]\psi=0$ | $A\to A,B\to B$<br>$E\to(\varepsilon^2-1)/2\bar{\lambda}^2$ |
| **Dirac-Woods-Saxon** | $\left[-\frac{d^2}{dr^2}-2\frac{\varepsilon B-\lambda^2/2}{1+e^{\lambda(r-R)}}-\frac{\varepsilon^2-1}{\bar{\lambda}^2}\right]\phi^+=0$ | $\left[-\frac{d^2}{dr^2}-\frac{2V_0}{1+e^{(r-R)/a}}-2E\right]\psi=0$ | $R\to R,a\to\lambda^{-1}$<br>$V_0\to\varepsilon B-\lambda^2/2$<br>$E\to(\varepsilon^2-1)/2\bar{\lambda}^2$ |

**Table 3**

| | $\phi_n^{+}(r)$ | Variables & Parameters |
|---|---|---|
| **Dirac-Coulomb** | $a_n(\lambda_n r)^{\gamma+1} e^{-\lambda_n r/2} L_n^{2\gamma+1}(\lambda_n r)$ | $\gamma=\sqrt{\kappa^2-\lambdabar^2 Z^2}\,,\lambda_n=-2Z\varepsilon_n/(\gamma+n+1)$ |
| **Dirac-Oscillator** | $a_n(\omega r)^{\kappa+1} e^{-\omega^2 r^2/2} L_n^{\kappa+1/2}(\omega^2 r^2)$ | —— |
| **Dirac-Morse** | $a_n z^{\nu_n} e^{-z/2} L_n^{2\nu_n}(z)$ | $z=\mu e^{-\lambda r}\,,\mu=2A/\lambda\,,\nu_n=B\varepsilon_n/\lambdabar A-n$ |
| **Dirac-Rosen-Morse I** | $a_n(1-z)^{\frac{\mu_n+\nu_n}{2}}(1+z)^{\frac{\mu_n-\nu_n}{2}} P_n^{(\mu_n+\nu_n,\mu_n-\nu_n)}(z)$ | $z=\tanh(\lambda r),\mu_n=A/\lambda-n,\nu_n=\varepsilon_n B/\mu_n\lambda^2$ |
| **Dirac-Eckart** | $a_n(z-1)^{\frac{\mu_n+\nu_n}{2}}(z+1)^{\frac{\mu_n-\nu_n}{2}} P_n^{(\mu_n+\nu_n,\mu_n-\nu_n)}(z)$ | $z=\coth(\lambda r),\mu_n=A/\lambda-n,\nu_n=\varepsilon_n B/\mu_n\lambda^2$ |
| **Dirac-Rosen-Morse II** | $a_n(z-1)^{\frac{\mu}{2}}(z+1)^{\frac{\nu}{2}} P_n^{(\mu-1/2,\nu-1/2)}(z)$ | $z=\cosh(\lambda r),\mu=(B-A)/\lambda\,,\nu=-(B+A)/\lambda$ |
| **Dirac-Scarf** | $a_n(1+z^2)^{-\frac{\mu}{2}} e^{-\nu\tan^{-1}(z)} P_n^{(-\mu-i\nu-1/2,-\mu+i\nu-1/2)}(iz)$ | $z=\sinh(\lambda r),\mu=A/\lambda\,,\nu=B/\lambda$ |
| **Dirac-Pöschl-Teller** | $a_n(1-z)^{\frac{\mu}{2}}(1+z)^{\frac{\nu}{2}} P_n^{(\mu-1/2,\nu-1/2)}(z)$ | $z=\cosh(2\lambda r),\mu=B/\lambda\,,\nu=A/\lambda$ |
| **Dirac-Woods-Saxon** | $a_n z^{\mu_n}(1-z)^{i\nu_n}\,{}_2F_1(\mu_n+i\nu_n,\mu_n+1+i\nu_n,2\mu_n+1;z)$ | $z=\left[1+e^{\lambda(r-R)}\right]^{-1},\mu_n^2=(1-\varepsilon_n^2)/\lambdabar^2\lambda^2\,,\nu_n^2=2\varepsilon_n B/\lambda^2-\mu_n^2-1$ |

**Table 4**

| **Potential Classes** | | $q(\rho)$ |
|---|---|---|
| **Dirac-Oscillator** | **Dirac-Coulomb** | $\rho^2$ |
| | **Dirac-Oscillator** (*reference*) | $\rho$ |
| | **Dirac-Morse** | $-(2/\lambda)\ln\rho$ |
| **Dirac-Rosen-Morse** | **Dirac-Rosen-Morse I** | $\lambda^{-1}\tanh^{-1}\left[\cosh(\lambda\rho)\right]$ |
| | **Dirac-Eckart** | $\lambda^{-1}\coth^{-1}\left[\cosh(\lambda\rho)\right]$ |
| | **Dirac-Rosen-Morse II** (*reference*) | $\rho$ |
| | **Dirac-Scarf** | $\lambda^{-1}\sinh^{-1}\left[-i\cosh(\lambda\rho)\right]$ |
| | **Dirac-Pöschl-Teller** | $\rho/2$ |